\documentclass[twocolumn,a4,showpacs,preprintnumbers,amsmath,amssymb]{revtex4-1}
\usepackage{epsfig}

\usepackage{mathptmx, courier, pifont}
\usepackage[scaled=0.92]{helvet}
\usepackage[T1]{fontenc}
\usepackage{textcomp}
\usepackage[pdftex,colorlinks=true]{hyperref}

\def\bra{\,<\!} \def\ket{\!>\,} \def\ack{\,|\,}
\begin{document}

\author{\large S. Jehangir$^{2}$}
\author{\large G.H. Bhat$^{1,3,4}$}
\author{\large J.A. Sheikh$^{4}$}
\author{\large S. Frauendorf$^5$}
\author{\large W. Li$^5$}
\author{\large R. Palit$^2$}

\affiliation{$^{1}$ Department of Physics, S.P. College,  Srinagar, Jammu and Kashmir, 190001, India}
\affiliation{$^2$Department of Nuclear and Atomic Physics, 
Tata Institute of Fundamental Research, Mumbai - 400005, India}
\affiliation{$^3$Cluster University Srinagar, Jammu and Kashmir,
  Srinagar, Goji Bagh, 190008, India}
\affiliation{$^4$Department of Physics, University of Kashmir,
  Hazratbal, Srinagar, 190006, India}
\affiliation{$^5$Physics Department, University of Notre Dame, Notre Dame, Indiana 46556, USA}


\title{Triaxial projected shell model study of $\gamma$-bands in
  atomic nuclei}

  \begin{abstract}

A systematic study of 
$\gamma$-bands observed in atomic nuclei is performed using the triaxial projected shell model (TPSM)
approach. The staggering phase between the even and odd spin members 
of the  $\gamma$-band  for most the nuclei
investigated in the present work is found to have even-I-below-odd-I, which in the framework of the collective model 
is considered as a signature of  $\gamma$-softness. It is observed that
out of twenty-three systems studied, only four nuclei, namely, $^{76}$Ge,
$^{112}$Ru, $^{170}$Er and $^{232}$Th depict staggering phase 
with odd-I-below-even-I, which  
is regarded as an indication of the static $\gamma$-deformation in the
collective model picture.
The inclusion of the quasiparticle excitations
in the framework of configuration mixing is shown to
 reverse the staggering phase from odd-I-down  to the even-I-down for all the studied nuclei,
 except for the aforementioned four nuclei. Furthermore, by fitting a
 collective Bohr Hamiltonian to the TPSM energies, the differences
 between the two models are delineated through a comparison of the transition
probabilities.
 \end{abstract}

\date{\today}


\maketitle
\section{Introduction}
Spontaneous breaking of rotational symmetry that leads to the
deformation of a quantum system in the intrinsic frame, has
played a pivotal role to unravel the underlying shapes and structures of
atomic nuclei \cite{SF01}. The properties of deformed nuclei are elucidated
by considering the ellipsoidal shape, which is conveniently 
parameterized in terms of axial and non-axial deformation parameters
of $\beta$ and $\gamma$. The majority of the deformed nuclei are
axially-symmetric ($\gamma=0$) with angular-momentum projection 
along the symmetry axis, $K$, a conserved quantum number with the
electromagnetic transition probabilities obeying the selection rules 
based on this quantum number \cite{nilson,BM75}.
There are also regions in the nuclear periodic table, 
referred to as transitional, where the axial symmetry is broken and
the non-axial degree of freedom plays an essential role 
to determine the properties of these nuclei. 

In the traditional picture, atomic nuclei may have either a localized minimum or a flat potential 
energy surface along the $\gamma-$ degree of
freedom, corresponding to $\gamma$-rigid and $\gamma$-soft
nuclei, respectively \cite{NV91,AS60,KM70,CB80}. 
How to distinguish between the two kinds of shapes
from the observable  properties has been  of considerable interest in nuclear physics 
for more than sixty years. To address the question properly one needs to
complement the potential  by inertial parameters
to estimate the spread  of the wave around the minimum, which is accomplished in various ways.
The  phenomenological Bohr Hamiltonian  \cite{BM75,MC11} assumes
irrotational-flow inertia and has the two limiting cases. The one
limiting case,
referred to as the Davydov-Filippov model \cite{AS58} describes a rigid triaxial shape, 
which corresponds to a deep potential minimum with respect to both $\beta$ and $\gamma$.
The second case, called as Wilets-Jean model \cite{LW56}, describes the completely
$\gamma$-soft limit and corresponds to a deep potential minimum with respect to $\beta$ and 
no $\gamma-$ dependence. 
Both limiting cases give rise to similar excitation spectrum for the
ground-state band \cite{AS58,LW56}. This holds as well for intermediate cases
studied in Ref. \cite{MC11}, where the author found that  the average energy of the $\gamma-$ band  
is insensitive to the rigidity of the $\gamma-$ degree of freedom.  
It is, therefore, impossible to distinguish between soft- and rigid- triaxiality from   
rich data of this kind that is available for most of the nuclei. 

The energy staggering in the $\gamma$-band, on the other hand, is sensitive to the softness of the 
$\gamma$-degree of freedom \cite{NV91}. For $\gamma$-soft nuclei (Wilets-Jean limit),
 the energies of the $\gamma-$band are bunched as $(2^+),(3^+,4^+),
(5^+,6^+)........$, which we refer to as "even-I-down",  and for
$\gamma$-rigid nuclei (Davydov-Filippov limit),
 the energy levels of the $\gamma$-band are
arranged as $(2^+,3^+),(4^+,5^+),(6^+,7^+)........$, which we shall call
as "odd-I-down". This implies that the sequence of
energy levels of the $\gamma$-band shall lead to opposite phase of the 
staggering parameter, defined below, in the two cases. More detailed
discussions on the subject can be found, e. g., in Refs. 
\cite{MC11,Frauendorf15}. The microscopic versions of the Bohr Hamiltonian,
pioneered in Ref. \cite{KM70} (for a review of recent work, see \cite{Frauendorf15}),
are based on a potential and inertial parameters, derived by applying
the adiabatic approximation to the time-dependent mean-field theory. Although not explicitly studied,
the same correlation between $\gamma$-softness and staggering of the $\gamma-$ band levels is found.
A similar pattern concerning the rigidity of the $\gamma-$ degree of freedom is obtained in
the framework of the interacting boson model  \cite{McCutchan07,Stefanescu07}. The correlation
between $\gamma-$ rigidity and staggering is discussed in detail in Ref. \cite{Frauendorf15}.
It has to be underlined at this point that correlation appears in
models that assume an adiabatic separation
between the collective quadrupole degrees of freedom and the quasiparticle excitations.

The present work investigates the $\gamma$-band
staggering in even-even nuclei using the microscopic triaxial projected shell
model (TPSM) approach \cite{JS99}, which is based on assumptions that
are different from above discussed collective models.
The TPSM assumes a {\it fixed}  triaxial deformation, and the coupling
to the quasiparticle excitations is taken into account using the
configuration mixing. The present study has been performed for all the nuclei in
the periodic table for which $\gamma$-bands are observed up to 
high angular momentum, for both even- and odd-spin members. It is
demonstrated that angular-momentum projection from the 
intrinsic triaxial vacuum  gives rise to odd-I-down $\gamma$-band staggering, which is expected
because the angular momentum  components projected from 0-quasiparticle state establish
a microscopic version of the Davydov-Filippov model of rigid-$\gamma$ motion. However, it is shown
that inclusion of quasiparticle excitations transforms the odd-I-down  
staggering into even-I-down staggering, associated with $\gamma$-softness,  for all 
even-even nuclei studied in the present
work, except for four nuclei of $^{76}$Ge, $^{112}$Ru, $^{170}$Er and $^{232}$Th. 

We have also investigated $\gamma$-bands using the Bohr
Hamiltonian by adjusting its parameters to reproduce the TPSM
transition energies. It is demonstrated that the $\gamma$-band
staggering obtained in this model is quite similar to that of the TPSM approach.
For $^{104,106,108}$Mo and $^{108,110,112}$Ru, we also evaluated
$B(E2)$ values for intra-band and inter-band transitions
in the two approaches. In comparison to
the TPSM results, a large reduction of the $B(E2, I_{\gamma}\rightarrow (I-1)_{ground}$ for $I>3$ and 
of the $B(E2, I_{\gamma}\rightarrow (I-2)_{ground}$ for $I>4$ values
is observed in the collective model.

The manuscript is organized in the following manner.
Section II contains  brief description of TPSM approach, more details on
the model  can be found in our earlier publications
\cite{GH14,bh15,GH08}. The results obtained from the TPSM
calculations on the nature of $\gamma$-bands are  analyzed in Section III. A.
In section III. B,  the TPSM  
transition probabilities are compared with the ones obtained using
the Bohr Hamiltonian and finally section IV contains some concluding remarks.

\section{Outline of the Triaxial Projected Shell Model Approach}
For even-even systems, the TPSM basis space is composed of projected 0-qp
state (or qp-vacuum $\ack\Phi\ket$), 2-proton, 2-neutron, and 4-qp
configurations, i.e.,
\begin{equation}
\begin{array}{r}
\hat P^I_{MK}\ack\Phi\ket;\\
~~\hat P^I_{MK}~a^\dagger_{p_1} a^\dagger_{p_2} \ack\Phi\ket;\\
~~\hat P^I_{MK}~a^\dagger_{n_1} a^\dagger_{n_2} \ack\Phi\ket;\\
~~\hat P^I_{MK}~a^\dagger_{p_1} a^\dagger_{p_2}
a^\dagger_{n_1} a^\dagger_{n_2} \ack\Phi\ket ,
\label{basis}
\end{array}
\end{equation}
where the three-dimensional angular-momentum operator \cite{RS80}
is given by
\begin{equation}
\hat P^I_{MK} = {2I+1 \over 8\pi^2} \int d\Omega\,
D^{I}_{MK}(\Omega)\, \hat R(\Omega),
\label{PD}
\end{equation}
with the rotation operator 
\begin{equation}
\hat R(\Omega)= e^{-i\alpha \hat J_z}e^{-i\beta \hat J_y}
e^{-i\gamma \hat J_z}.\label{rotop}
\end{equation}
Here, $''\Omega''$ represents a set of Euler angles 
($\alpha, \gamma = [0,2\pi],\, \beta= [0, \pi]$) and the 
$\hat{J}^{'}s$ are angular-momentum operators.
The triaxial vacuum
configuration is a superposition of K-configurations and it can be
easily shown that only even-K values are permitted due to symmetry
requirement \cite{YK00}. The projected bands from the vacuum state 
with K=0, 2 and 4 in the $D$-matrix result into ground-,
$\gamma$- and $\gamma\gamma$-bands, respectively. For
two-quasiparticle states, both even- and odd-K values are permitted,
depending on the nature of the quasiparticles. For a two quasiparticle
configuration formed from normal and time-reversed states, only even-K are
permitted. However, with both the states either normal or
time-reversed, odd-K values are allowed from symmetry
considerations. 

The projected basis constructed above are then employed to diagonalize
the shell model Hamiltonian. As in the earlier TPSM calculations, we 
use the pairing plus
quadrupole-quadrupole Hamiltonian$:$ 
\begin{equation}
\hat H = \hat H_0 - {1 \over 2} \chi \sum_\mu \hat Q^\dagger_\mu
\hat Q^{}_\mu - G_M \hat P^\dagger \hat P - G_Q \sum_\mu \hat
P^\dagger_\mu\hat P^{}_\mu,
\label{hamham}
\end{equation}
with the last term in (\ref{hamham}) being the quadrupole-pairing
force.  The corresponding triaxial Nilsson mean-field Hamiltonian,
which  is obtained by using the Hartree-Fock-Bogoliubov (HFB)
approximation, is given by
\begin{equation}
\hat H_N = \hat H_0 - {2 \over 3}\hbar\omega\left\{\epsilon\hat Q_0
+\epsilon'{{\hat Q_{+2}+\hat Q_{-2}}\over\sqrt{2}}\right\}.
\label{nilsson}
\end{equation}
Here $\hat H_0$ is the spherical single-particle Hamiltonian, which
contains a proper spin-orbit force \cite{Ni69}. The interaction
strengths are taken as follows:  The $QQ$-force strength $\chi$ is
adjusted such that the physical quadrupole deformation $\epsilon$ is
obtained as a result of the self-consistent mean-field HFB
calculations.  The monopole pairing strength $G_M$ is of
the standard form
\begin{equation}
G_{M} = (G_{1}\mp G_{2}\frac{N-Z}{A})\frac{1}{A} \,(\rm{MeV}),
\label{gmpairing}
\end{equation}
 where $- (+)$ is neutron (proton).  
In the present calculation, we use $G_1$ and $G_2$,
which approximately reproduce the observed odd-even mass difference
in the region under investigation. This choice of $G_M$ is appropriate for the
single-particle space employed in the model, where three major
shells are used for each type of nucleons. The quadrupole pairing strength $G_Q$ is
assumed to be proportional to $G_M$, and the proportionality
constant being fixed as 0.16.  These interaction strengths are
consistent with those used in our earlier studies
\cite{GH14,bh15,GH08}.

The projection formalism outlined above
can be transformed into a diagonalization problem following the
Hill-Wheeler approach, i.e., 
\begin{eqnarray}
\sum_{K',\kappa'}( \bra \Phi_{\kappa}|&\hat H& \hat P^{I}_{KK'}|\Phi_{\kappa'}\ket
\\\nonumber&-&E_l \bra \Phi_{\kappa}|\hat P^{I}_{KK'}|\Phi_{\kappa^{\prime}}\ket)f^I_{K'\kappa'} = 0~~~,
\label{HW}
\end{eqnarray}
where $f^I_{K\kappa}$ are the variational coefficients.
The projected wavefunction in terms of these coefficients can be written as
\begin{equation}
\psi_{IM} = \sum_{K,\kappa}~f^I_{K\kappa}~\hat P^{I}_{MK}| 
~ \Phi_{\kappa} \ket.
\label{17}
\end{equation}
In the above equations, the symbol $\kappa$ represents the basis
states of Eq.~(\ref{basis}). These wavefunctions are used to 
calculate the electromagnetic transition probabilities.   The reduced electric quadrupole transition probability $B(E2)$ from an initial state 
$( \sigma_i , I_i) $ to a final state $(\sigma_f, I_f)$ is given by \cite {su94}
\begin{equation}
B(E2,I_i \rightarrow I_f) = {\frac {e^2} {2 I_i + 1}} 
| \bra \sigma_f , I_f || \hat Q_2 || \sigma_i , I_i\ket |^2 .
\label{BE22}
\end{equation}
As in our earlier publications \cite{JG11,GJ12,JG12,GH14,bh15}, we have used the effective charges
of 1.5e for protons and 0.5e for neutrons. The effective charges are
employed instead of the bare charges as core is used in the TPSM and
valance particles occupy only three major oscillator shells. 
The reduced magnetic dipole transition probability
$B(M1)$ is computed through
\begin{equation}
B(M1,I_i \rightarrow I_f) = {\frac {\mu_N^2} {2I_i + 1}} | \bra \sigma_f , I_f || \hat{\mathcal M}_1 ||
\sigma_i , I_i \ket | ^2 , 
\label{BM11}
\end{equation}
where the magnetic dipole operator is defined as  
\begin{equation}
\hat {\mathcal {M}}_{1}^\tau = g_l^\tau \hat j^\tau + (g_s^\tau - g_l^\tau) \hat s^\tau . 
\end{equation}  
Here, $\tau$ is either $\nu$ or $\pi$, and $g_l$ and $g_s$ are the orbital and the spin gyromagnetic factors, 
respectively. 
In the calculations
we use for $g_l$ the free values and for $g_s$ the free values 
damped by a 0.85 factor, i.e.,
\begin{eqnarray}
&&g_l^\pi = 1, ~~~ 
g_l^\nu = 0, ~~~   \nonumber\\
&&g_s^\pi =  5.586 \times 0.85,\nonumber\\
&& g_s^\nu = -3.826 \times 0.85.
\end{eqnarray}
The reduced matrix element of an operator $\hat {\mathcal {O}}$ ($\hat {\mathcal {O}}$ is either
$\hat {Q}$ or $\hat {\mathcal {M}}$) can be expressed as
\begin{eqnarray}
&&\bra \sigma_f , I_f || \hat {\mathcal {O}}_L || \sigma_i , I_i\ket 
\nonumber \\ & =& 
\sum_{\kappa_i , \kappa_f} {a_{ \kappa_i}^{\sigma_i}} {a_{ \kappa_f}^{\sigma_f}}
 \sum_{M_i , M_f , M} (-)^{I_f - M_f}  \nonumber \\&\times&
\left( \begin{array}{ccc}
 I_f & L & I_i \\
-M_f & M & M_i 
\end{array} \right) \nonumber \\
 &\times&  \bra \Phi_{\kappa_f} | {\hat{P}^{I_f}}_{K_{\kappa_f} M_f} \hat {\mathcal {O}}_{LM}
\hat{P}^{I_i}_{K_{\kappa_i} M_i} | \Phi_{\kappa_i} \ket  \nonumber \\
&=& 2 \sum_{\kappa_i , \kappa_f} {a_{ \kappa_i}^{\sigma_i}} {a_{ \kappa_f}^{\sigma_f}}\sum_{M' , M''} (-)^{I_f - K_{\kappa_f}} {(2I_f + 1)}^{-1}\nonumber\\
&\times&   \left( \begin{array}{ccc}
 I_f & L & I_i \\
-K_{\kappa_f} & M' & M'' 
\end{array} \right)~\int d\Omega \,D^{I_i}_{M''K_{\kappa_i}}(\Omega)\nonumber\\
&\times&\bra \Phi_{\kappa_f}|\hat {\mathcal {O}}_{LM'}\hat R(\Omega)|\Phi_{\kappa_i}\ket.
\label{beeee2}
\end{eqnarray}
In the above expression, the symbol $(~~~~~~)$ denotes a 3j-coefficient.

\begin{table}
\caption{Parameters used in calculations}
\begin{tabular}{|cccc|}
  \hline  Nuclei   &Configuration Space & G$_1$ &G$_2$ \\\hline
  
  $^{154,156}$Gd       &N$_\nu$ = 4, 5, 6      &  21.24   &13.86\\
  $^{156,158,160,162}$Dy    &N$_\pi$ = 3, 4, 5      &          &\\
  $^{164,166,168,170}$Er    &                      &          &\\
  $^{180}$Hf &                     &          &\\
    \hline
$^{232}$Th,$^{238}$U            &N$_\nu$ = 5, 6, 7    & 16.80    &12.80\\
  &N$_\pi$ = 4, 5, 6    &          &\\
  \hline
$^{104,106,108}$Mo&     N$_\nu$ = 3, 4, 5     &22.68     &16.22\\
  $^{108,110,112,114}$Ru&  N$_\pi$ = 2, 3, 4     &          &\\
  \hline
$^{76}$Ge,  $^{76,78}$Se & N$_\nu$= 2, 3, 4  &20.82     &13.58\\
  & N$_\pi$=2, 3, 4 &&\\
  \hline
 \end{tabular}\label{tab1}
\end{table}
\begin{table*}\scriptsize{
\caption{Axial and triaxial quadrupole deformation parameters
$\epsilon$ and $\epsilon'$  employed in the TPSM calculation. Axial deformations are taken from \cite{Raman}
and nonaxial deformations are chosen in such a way that band heads
of the $\gamma-$ bands are reproduced. In this table, we also provide
the ratios, ${\frac{E(2^+_2)} {E(2^+_1)}}$, using both experimental and TPSM
energies.}
\begin{tabular}{|cccccccccccccccccccccccc|}
\hline                   & $^{154}$Gd &  $^{156}$Gd    &  $^{156}$Dy  &$^{158}$Dy  & $^{160}$Dy& $^{162}$Dy  & $^{164}$Er  &$^{166}$Er & $^{168}$Er  &$^{170}$Er  &$^{180}$Hf &  $^{232}$Th & $^{238}$U & $^{104}$Mo & $^{106}$Mo & $^{108}$Mo & $^{108}$Ru & $^{110}$Ru & $^{112}$Ru & $^{114}$Ru &$^{76}$Ge&$^{76}$Se&$^{78}$Se\\
\hline   $\epsilon$      &0.300      &  0.341       &  0.278      & 0.260     & 0.270    &  0.280    &  0.317     &    0.325  &   0.321    &   0.319   &   0.195  &    0.248  &  0.210   &    0.320  &    0.310  &   0.294  &    0.280  &  0.290    &    0.289  &    0.250  &   0.200  & 0.260  &   0.256  \\
\hline $\epsilon'$       &0.100      &  0.100       &  0.105      & 0.110     & 0.110    &  0.120    &  0.120     &    0.126  &   0.130    &   0.110   &   0.090  &    0.085  &  0.085   &    0.130  &    0.110  &   0.140  &    0.150  &  0.150    &    0.130  &    0.080  &   0.160  &  0.155 &   0.150\\
\hline $\gamma$          & 18.4      &  16.3        & 20.7        & 22.9      & 22.1     & 23.2      & 20.7       &   21.2    &   22.1     & 19.0      & 24.7     &    18.9   & 22.0     &  22.1     & 19.5      &  25.4    &    28.2   &  27.3     &  24.2     &    17.7    & 38.6    & 30.8   &  30.2\\
\hline ${\lbrack\frac{E(2^+_2)}{E(2^+_1)}\rbrack}_{Expt.}$&8.0         &12.9          &6.5          &9.6        &11.1      &9.7         &10.4        &   9.7     &  10.3          &11.8       &12.9      &    15.9   &21.2      &4.2        &4.1        &3.0       &     2.9   &2.5        &2.2        &    2.1     &1.9      &2.2     &2.1\\
\hline ${\lbrack\frac{E(2^+_2)}{E(2^+_1)}\rbrack}_{TPSM}$ &10.6        &12.1          &9.7          &9.5        &12.9      &11.9        &9.4        &  10.5     &  11.7          &13.4       &15.4      &    15.3   &20.7      &5.3        &5.1        &3.6       &     3.5   &3.4        &3.2        &    2.4     &2.1      &2.2     &2.2\\

\hline
\end{tabular}\label{tab1}}
\end{table*}
\begin{figure}[htb]
 \centerline{\includegraphics[trim=0cm 0cm 0cm
0cm,width=0.45\textwidth,clip]{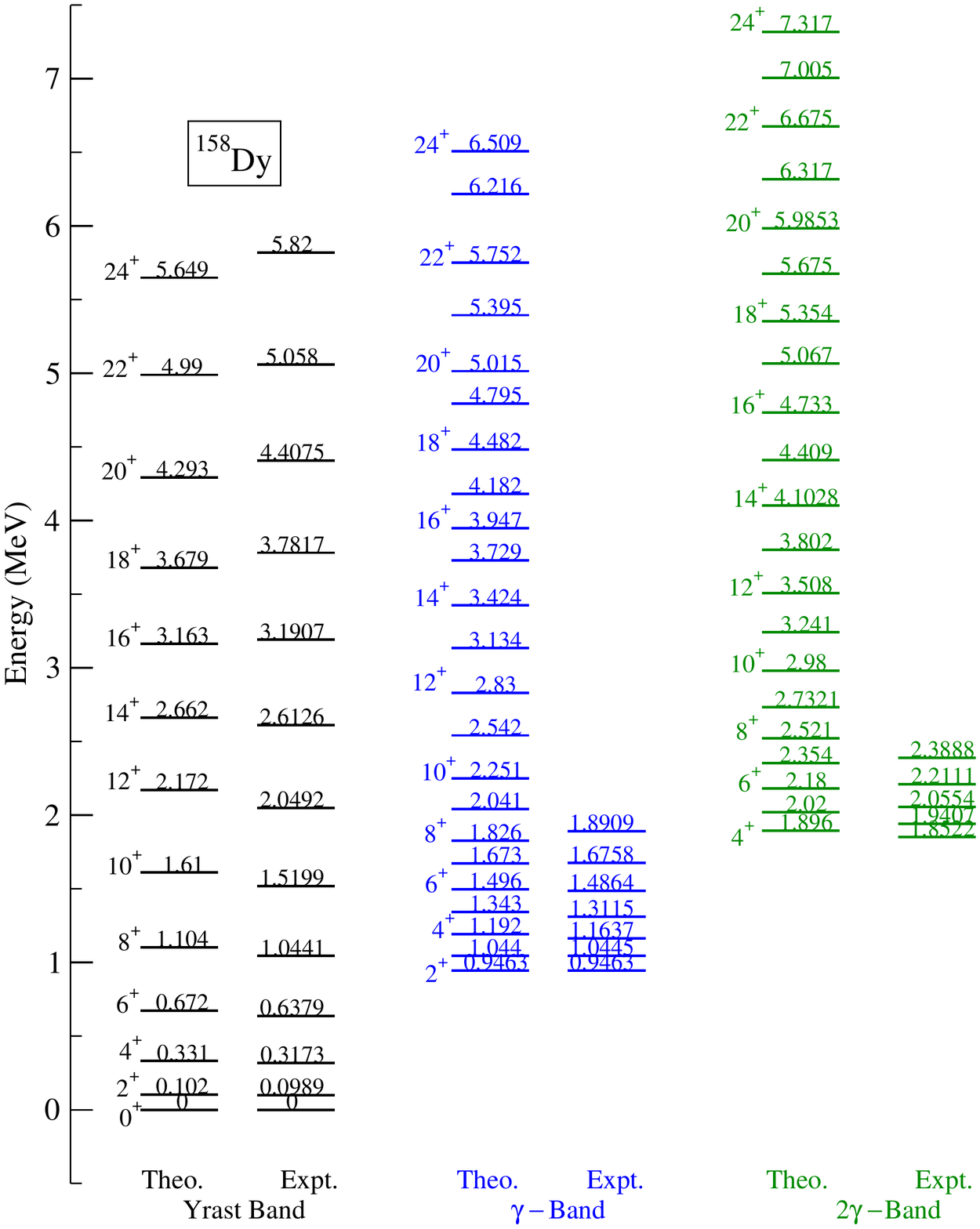}}
\caption{(Color online) Comparison of the TPSM energies after configuration mixing
with the available experimental data for $^{158}$Dy. Data is
taken from \cite{ML80,HE81}. }\label{fig0}
\end{figure}
\begin{figure}[htb]
 \centerline{\includegraphics[trim=0cm 0cm 0cm
0cm,width=0.5\textwidth,clip]{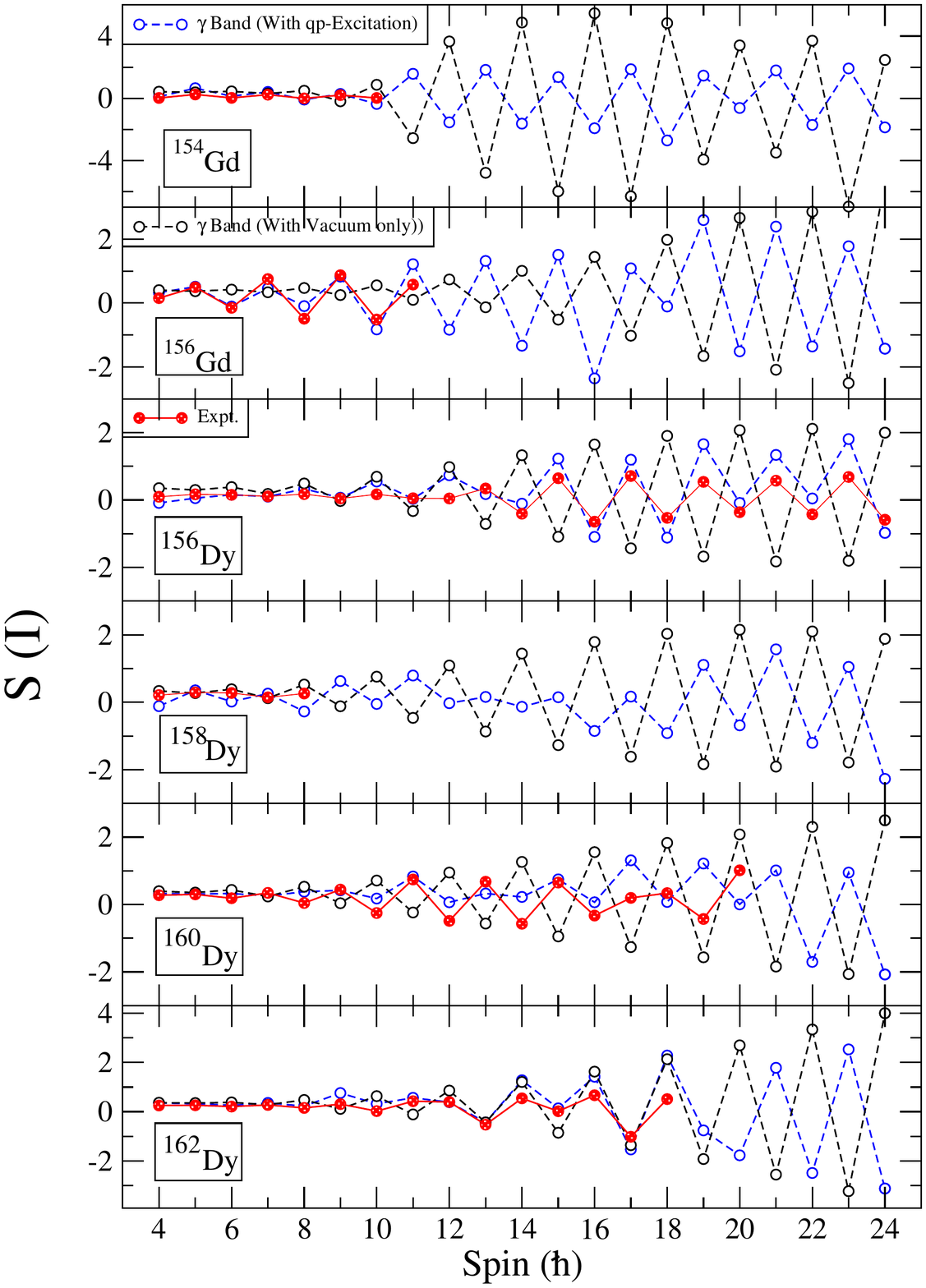}}
\caption{(Color online)  Comparison of observed and TPSM calculated staggering
parameter Eq. (\ref{eq:staggering}) for the
$\gamma$-band with and without quasiparticle excitations for $^{156,158}$Gd and $^{156,158,160,162}$Dy, nuclei. Data is taken from
Refs.~\cite{CD00,CF96,156dy,158dy,160dy,162dy,162dya}. }\label{fig:sGd}
\end{figure}

\begin{figure}[htb]
 \centerline{\includegraphics[trim=0cm 0cm 0cm
0cm,width=0.5\textwidth,clip]{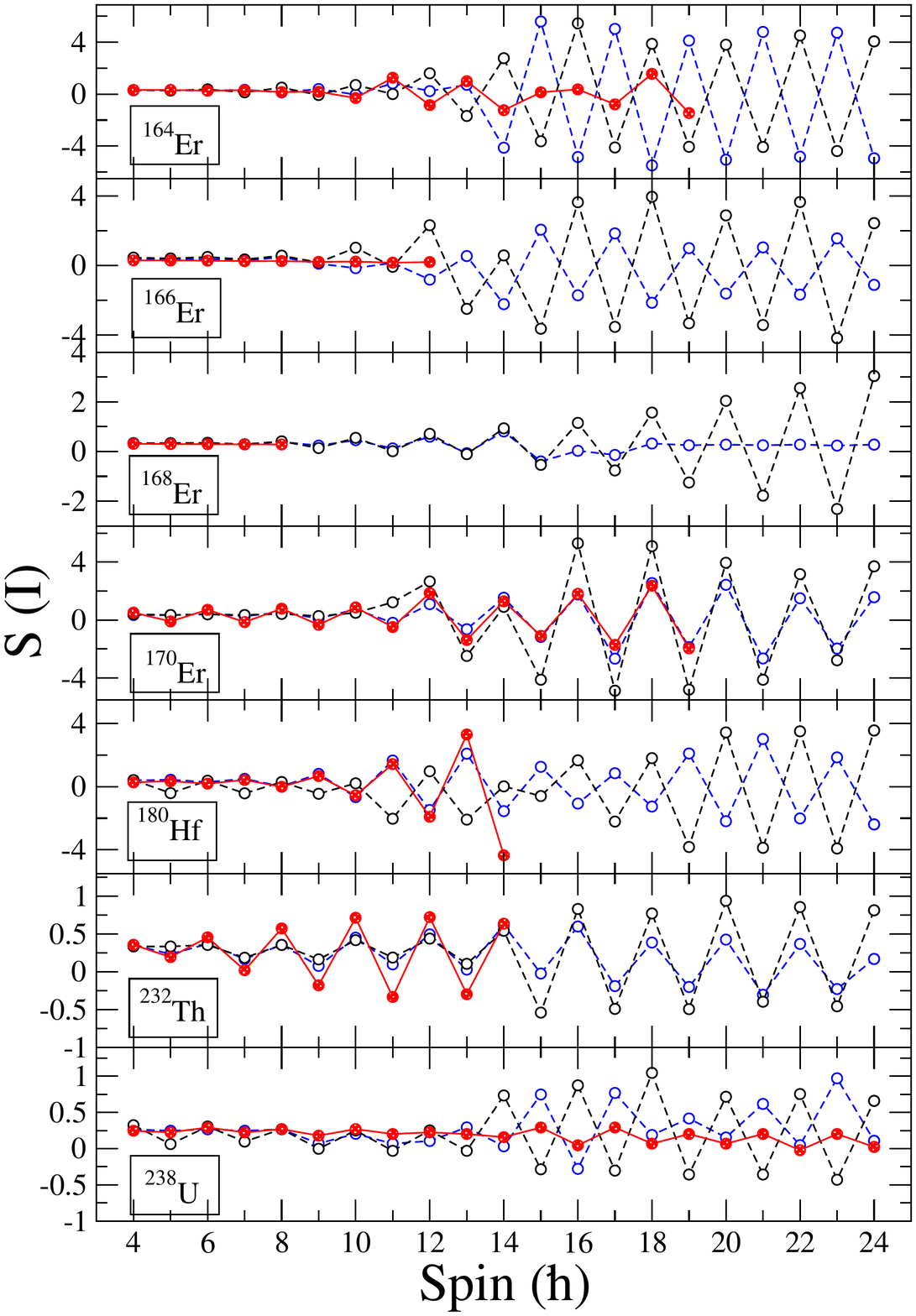}}
\caption{(Color online)  Comparison of observed and TPSM calculated staggering
parameter Eq. (\ref{eq:staggering}) for the
$\gamma$-band with and without quasiparticle excitations for  $^{164,166,168,170}$Er, $^{180}$Hf,  $^{232}$Th and $^{238}$U nuclei. Data is taken from
Refs.~\cite{TH98,GD10,180hf,180hg,232Th,238U}.}\label{fig:sEr}
\end{figure}

\begin{figure}[htb]
 \centerline{\includegraphics[trim=0cm 0cm 0cm
0cm,width=0.5\textwidth,clip]{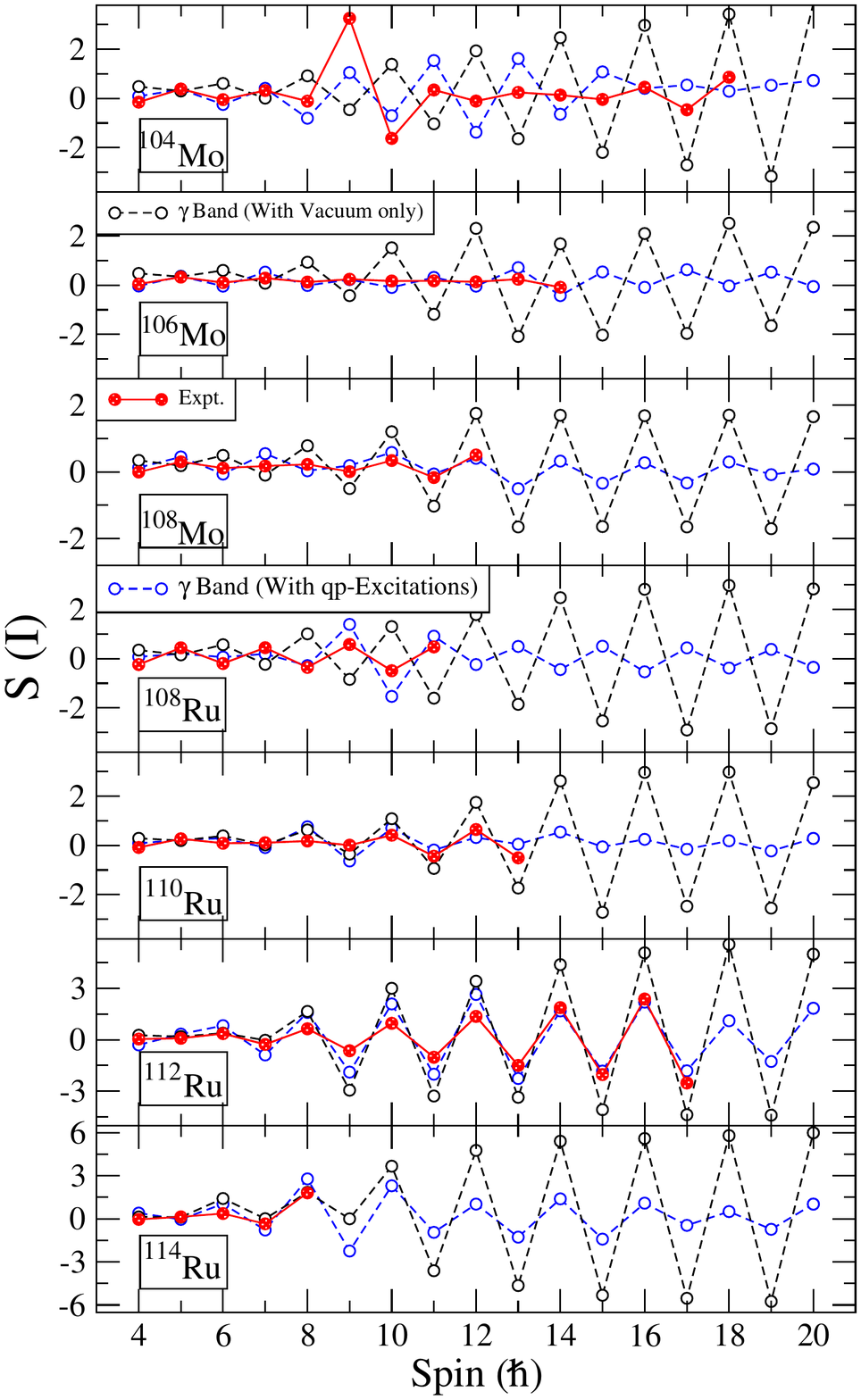}}
\caption{(Color online) Comparison of observed and TPSM calculated staggering
parameter Eq. (\ref{eq:staggering}) for the
$\gamma$-band with and without quasiparticle excitations for $^{104,106,108}$Mo and $^{108,110,112,114}$Ru nuclei.  Data is taken from
Refs.~\cite{AG96,HY04,LY01,CH04,JZ03}.  }\label{fig:sMo}
\end{figure}

\begin{figure}[htb]
 \centerline{\includegraphics[trim=0cm 0cm 0cm
0cm,width=0.5\textwidth,clip]{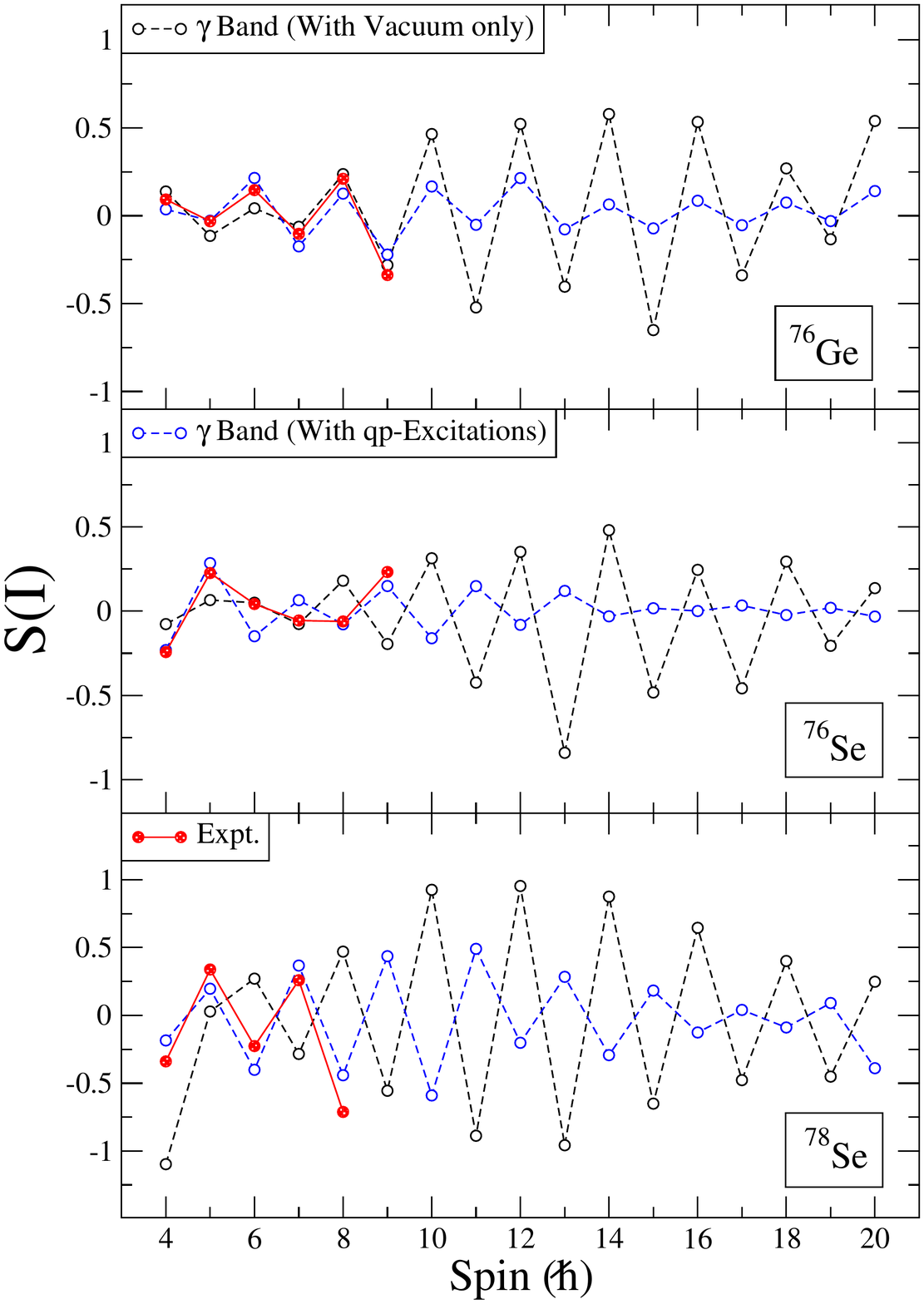}}
\caption{(Color online) Comparison of observed and TPSM calculated staggering
parameter Eq. (\ref{eq:staggering}) for the
$\gamma$-bands in $^{76}$Ge and $^{76,78}$Se nuclei. Data is taken from
Refs.~(\cite{YT13,DA76,DA78} }\label{fig:sGeSe}
\end{figure}

\begin{figure}[htb]
 \centerline{\includegraphics[trim=0cm 0cm 0cm
0cm,width=0.5\textwidth,clip]{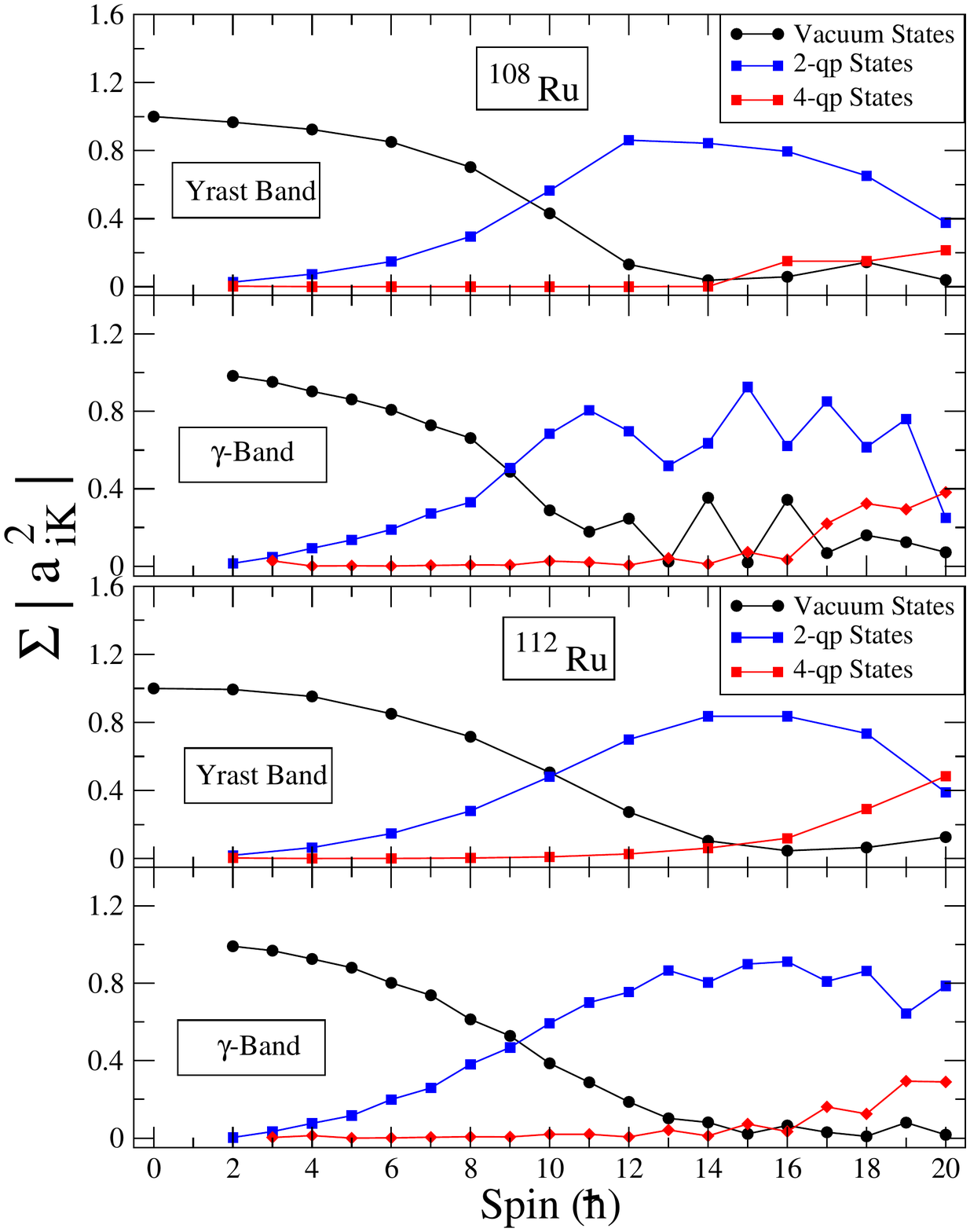}}
\caption{(Color online) Probability of various projected K-configurations in the
wavefunctions of the band structures after diagonalization are plotted
for the $^{108,112}$Ru isotopes.}\label{fig:pRu}
\end{figure}
\begin{figure}[htb]
 \centerline{\includegraphics[trim=0cm 0cm 0cm
0cm,width=0.5\textwidth,clip]{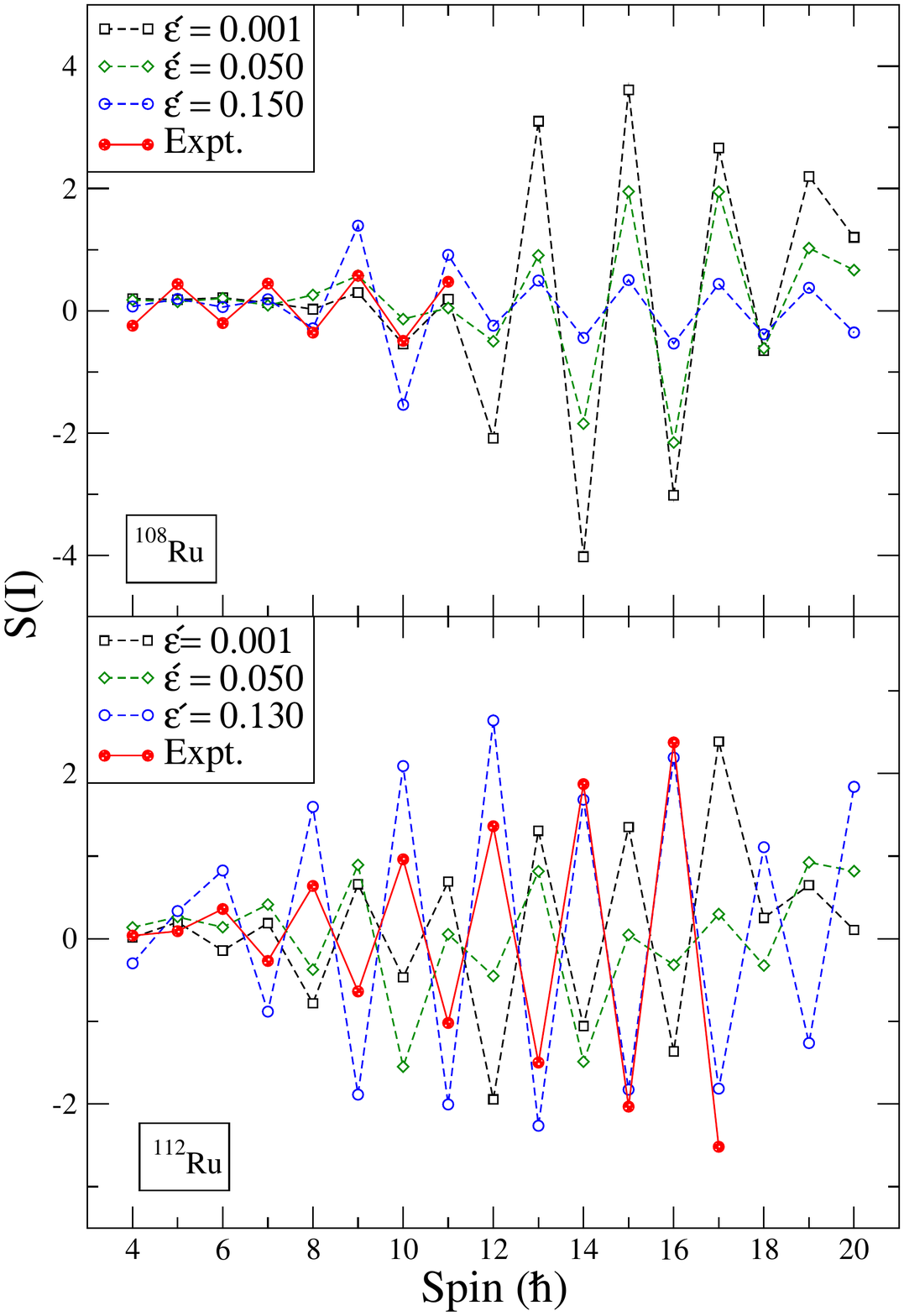}}
\caption{(Color online) Comparison of observed and TPSM calculated staggering
parameter Eq. (\ref{eq:staggering}) for  the
$\gamma$-bands in $^{108,112}$Ru isotopes for the different values of $\epsilon'$. }\label{fig:sRu}
\end{figure}

\section{Results and Discussions}

It has been demonstrated in several previous studies
\cite{JG11,GJ12,GH08,JG12,GH14,bh15,SJ18,JS16} that TPSM approach reproduces the high-spin properties of 
deformed and transitional nuclei quite accurately. As an illustrative
example,  TPSM calculated energies of the yrast-,$\gamma-$ and
$\gamma\gamma-$ bands of $^{158}$Dy nucleus are displayed in
Fig.~\ref{fig0} along with the known experimental energies. It is 
evident from the results that TPSM approach reproduces the
experimental data quite reasonably. In particular, it is noted that
predicted $\gamma$-band transition energies agree quite well with the
known experimental values. In the following, we now
address the main topic of the manuscript : to investigate the 
correlation between the nature of the $\gamma-$ deformation and 
the staggering phase of the $\gamma$-bands.

\subsection{Staggering of the $\gamma$-bands}\label{s:stag}

A systematic investigation of the $\gamma$-bands using TPSM framework has been performed for 
even-even nuclei listed in Table I. For the chosen twenty-three nuclei,
$\gamma$-bands for both even- and odd-spin signatures are observed up to
quite high-spin and it is possible to investigate the odd-even
staggering as a function of the spin. This table also provides the information on the
configuration space and pairing strengths used in different regions. Deformation
parameters of $\epsilon$ and $\epsilon'$ employed in the TPSM 
calculations are provided in Table II. The axial deformation parameter,  $\epsilon$
is either fixed such that the observed quadrupole moment of the first
excited state is reproduced or from other theoretical studies. The triaxiality
parameter, $\epsilon'$ is determined so that the band-head of
the $\gamma$-band is reproduced, a prescription adopted in most of our
earlier works \cite{JG12,GH14,bh15,SJ18,JS16}.
There are other ways of choosing this
parameter. One can use the tidal wave version of the cranking model \cite{FR11}, which has been applied to $^{156}$Dy \cite{SJ18} .
 A related  approach is  to look for the potential energy minimum
 with respect to $\epsilon'$ as a generator coordinate. It has
been shown in our earlier work \cite{GH14} that this leads to a similar
nonaxial deformation value as that deduced through fixing the
band-head energy of the $\gamma$-band.                 

As alluded to in the introduction, in the framework of collective
models with quadrupole degrees freedom,  
  the staggering parameter, defined as
\begin{equation}\label{eq:staggering}
S(I)= \frac{[E(I)-E(I-1)]-[E(I-1)-E(I-2)]}{E(2^{+}_1)},
\end{equation}
is strongly correlated with the rigidity of the triaxial shape : the odd-I-down pattern indicates the concentration of the collective 
wave function around a finite $\gamma-$ value (static triaxiality),
whereas the even-I-down pattern points to a spread of the wave function over the whole range of $\gamma$ (dynamic triaxiality).
The correlation is reviewed in Ref. \cite{Frauendorf15}, where the relevant literature is cited. 

In Figs.~\ref{fig:sGd} - \ref{fig:sMo}, the TPSM results  for $S(I)$
are plotted for twenty  nuclei in different regions of the 
nuclear chart, which have been selected as
$\gamma$-bands, both for even- and odd-spin signatures, 
have been observed  up to quite high spin. In
Figs.~\ref{fig:sGd} and \ref{fig:sEr} the results of $S(I)$ are depicted for $^{154,156}$Gd, $^{156,158,160,162}$Dy, $^{164,166,168,170}$Er, $^{180}$Hf,  $^{232}$Th
and $^{238}$U. The figures compare the TPSM calculations that are obtained
from the projection of zero-quasiparticle state only 
with the full TPSM calculations that include all the projected
quasiparticle configurations, listed in Eq.~\ref{basis}. 
The restricted  TPSM calculations represent a microscopic version of
the Davydov-Filippov model :
the different $K$ components for given $I$ are projected from one and the same intrinsic zero-quasiparticle state and the
resulting matrix of the TPSM Hamiltonian diagonalized.  Therefore, it is expected that $S(I)$ depicts the odd-I-down pattern of the rigid rotor,
which is borne out of the calculations.

However, Figs.~\ref{fig:sGd} and \ref{fig:sEr} demonstrate that the phase of the staggering changes
to even-I-down when the quasiparticle excitation in the configuration
mixing framework are taken into account for all the nuclei,
except for $^{170}$Er and $^{232}$Th. For the case of
$^{162}$Dy, the staggering phase in the spin regime from I=12 to 18 
is same with and without quasiparticle excitations, however, for
higher spin states the phase reverses. The TPSM values correlate well
with the available experimental staggering 
pattern, which are also plotted in the figures. The only exceptional
case is $^{238}$U, for which the difference between the calculated and the 
experimental staggering appears somewhat larger. However, it may be noted that
the magnitude of the staggering as compared to other nuclei  in
Fig.~\ref{fig:sEr} is quite small for this system.

The staggering parameters for Mo- and  Ru-isotopes are depicted in
Fig.~\ref{fig:sMo}. For $^{104,106}$Mo, the phase of staggering changes when the
quasiparticle excitations are taken into account.
For $^{108}$Mo, inclusion of the quasiparticle excitation reverses the staggering phase up to  
$I=8$, but above this spin the staggering remains odd-I-down. In the
case of $^{108}$Ru, inclusion of the quasiparticle excitations reverses the staggering  phase. 
For  $^{110}$Ru,  the magnitude of 
the staggering phase is small for low spin and becomes odd-I-down for high spin. 
The nuclide  $^{112}$Ru shows well pronounced odd-I-down staggering, which
is not modified when including the quasiparticle excitations. 
The case of  $^{114}$Ru is similar to  $^{110}$Ru, only the amplitude of the staggering is smaller.
As seen in Fig. ~\ref{fig:sMo}, the TPSM  results correlate impressively well with the experimental $S(I)$ values.
It is noted that in $^{108}$Mo and  $^{110}$Ru, the staggering changes from weak even-I-up to odd-I-down around I=8.
In  $^{112,114}$Ru, the odd-I-down staggering sets-in at I=6, which is
reproduced by the TPSM results.

In the mass $~$80 region, $\gamma$-bands have also been observed in
some nuclei up to high spin. In Ref. \cite{GH14},  the isotopes $^{76-82}$Se and $^{70-80}$Ge
have been investigated in the framework TPSM approach. The staggering parameter, $S(I)$ has been 
found with even-I-down for all nuclides except   $^{76}$Ge, which has odd-I-down. The TPSM results
 for $^{76} $Ge and $^{76,78}$Se agree well with the experimental $S(I)$  as shown in Fig.~\ref{fig:sGeSe}.
 Based on the odd-I-down pattern \cite{YT13} and $E2$ reduced
 transition matrix elements measured in Coulomb excitation experiment \cite{AD19},   $^{76}$Ge has been featured as 
a rare example of a $\gamma$-rigid nucleus.

\begin{figure}[h]
 \centerline{\includegraphics[trim=0cm 0cm 0cm
0cm,width=0.5\textwidth,clip] {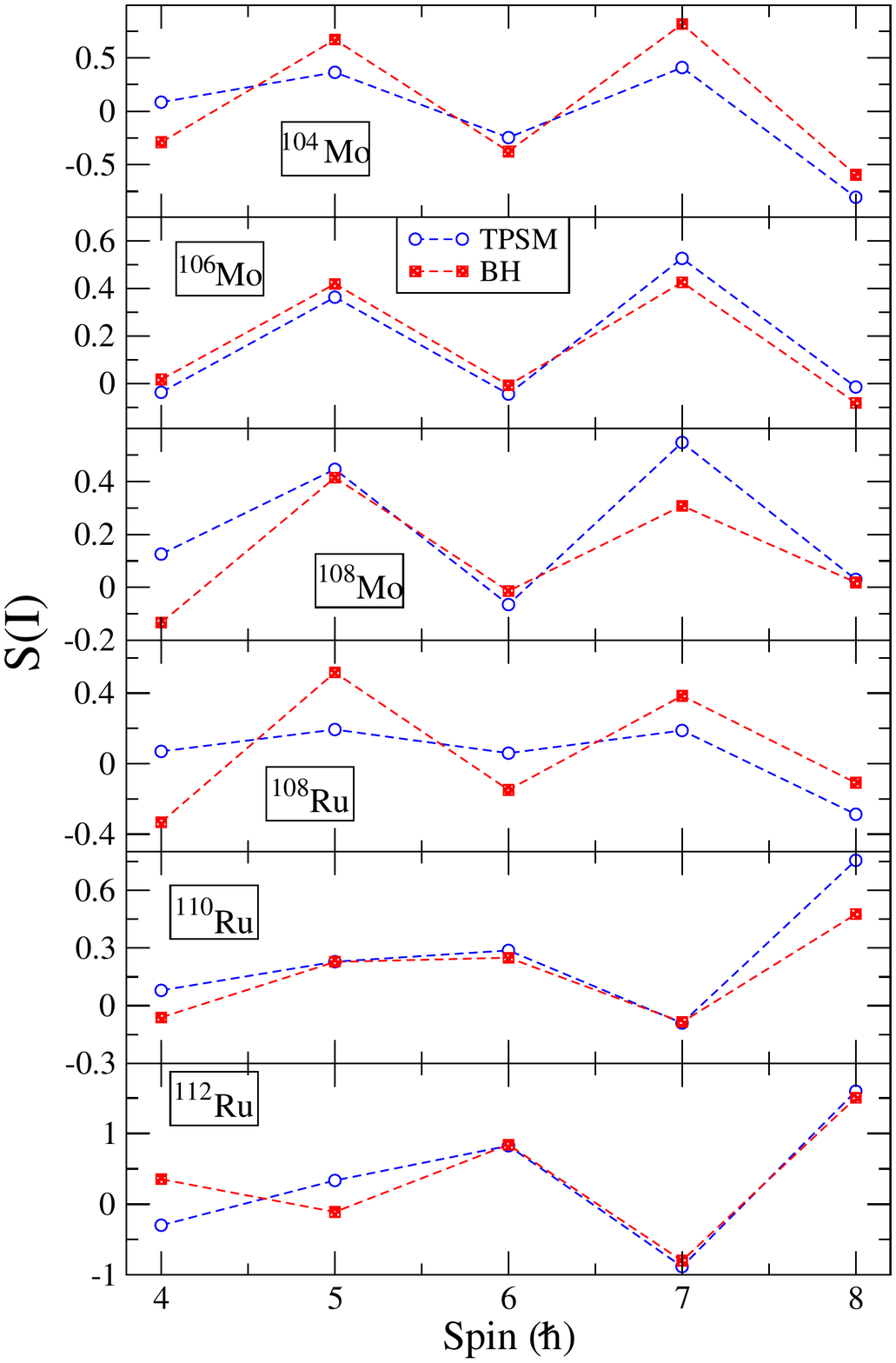}}
\caption{(Color online) Comparison of the staggering parameter calculated by means of the TPSM (blue circles) and
the BH (red squares).  }\label{fig:MoRusTPSMvsBH}
\end{figure}
\begin{figure}[htb]
 \centerline{\includegraphics[trim=0cm 0cm 0cm
0cm,width=0.5\textwidth,clip]{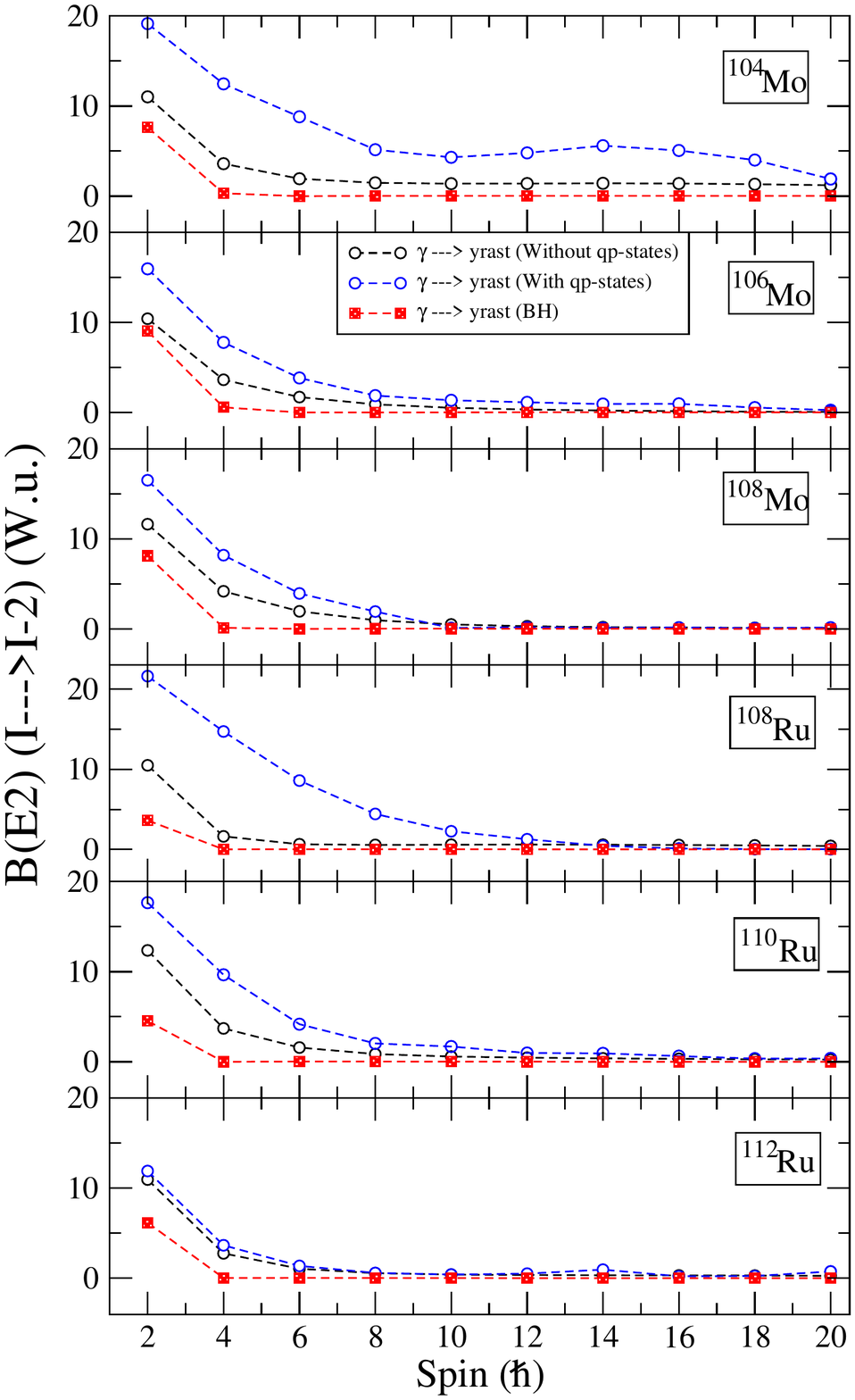}}
\caption{(Color online) Reduced E2  probabilities for $\Delta I=2$ transitions from
the $\gamma$ band to the ground band. The blue and black circles show the  TPSM values with and without 
coupling to the quasiparticle excitations, respectively.  The red squares 
show the BH values.  }\label{fig:Mo_Ru_BE2_I_I_2_g_y}
\end{figure}
\begin{figure}[t]
 \centerline{\includegraphics[trim=0cm 0cm 0cm
0cm,width=0.5\textwidth,clip]{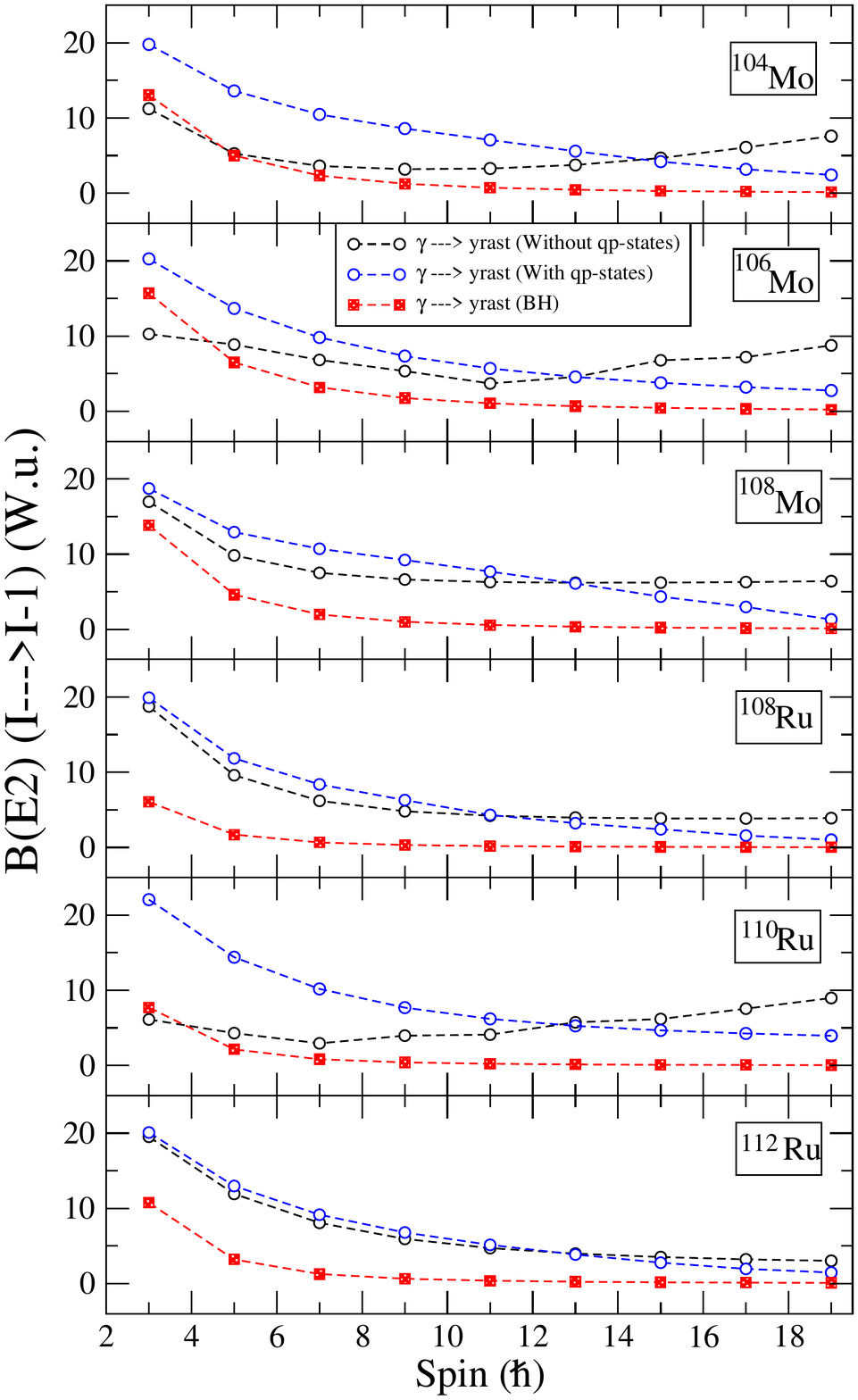}}
\caption{(Color online) Reduced E2  probabilities for $\Delta I=1$ transitions from
the $\gamma$ band to the ground band. The blue and black circles show the  TPSM values with and without 
coupling to the quasiparticle excitations, respectively.  The red squares 
show the BH values.  }\label{fig:Mo_Ru_BE2_I_I_1_g_y}
\end{figure}

\begin{figure}[t]
 \centerline{\includegraphics[trim=0cm 0cm 0cm
0cm,width=0.5\textwidth,clip]{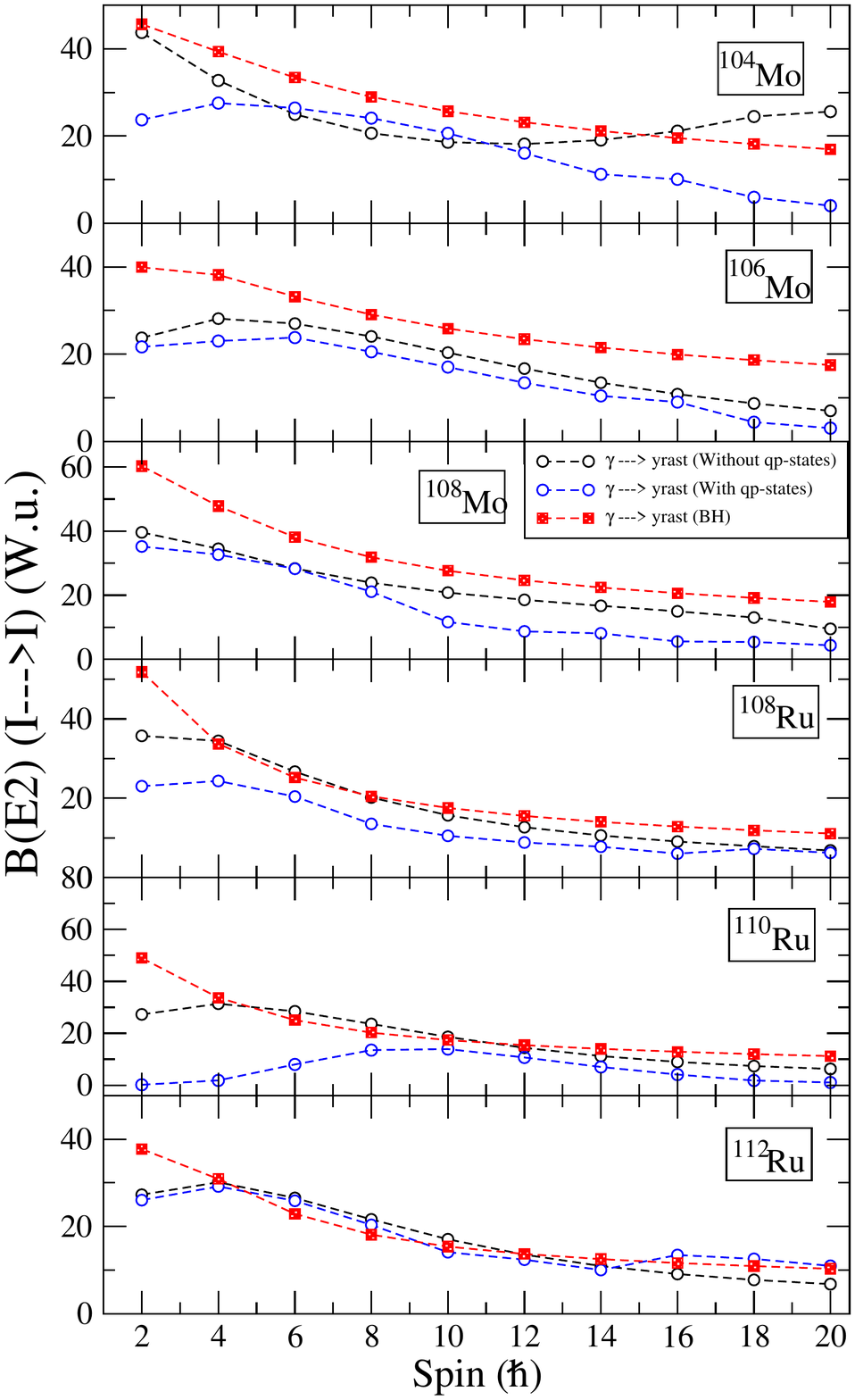}}
\caption{(Color online)Reduced E2  probabilities for $\Delta I=0$ transitions from
the $\gamma$ band to the ground band. The blue and black circles show the  TPSM values with and without 
coupling to the quasiparticle excitations, respectively.  The red squares 
show the BH values. Except for $^{112}$Ru the values for the lowest spins are out of the frame. }\label{fig:Mo_Ru_BE2_I_I_0_g_y}
\end{figure}
\begin{figure}[h] 
 \centerline{\includegraphics[trim=0cm 0cm 0cm
0cm,width=0.5\textwidth,clip]{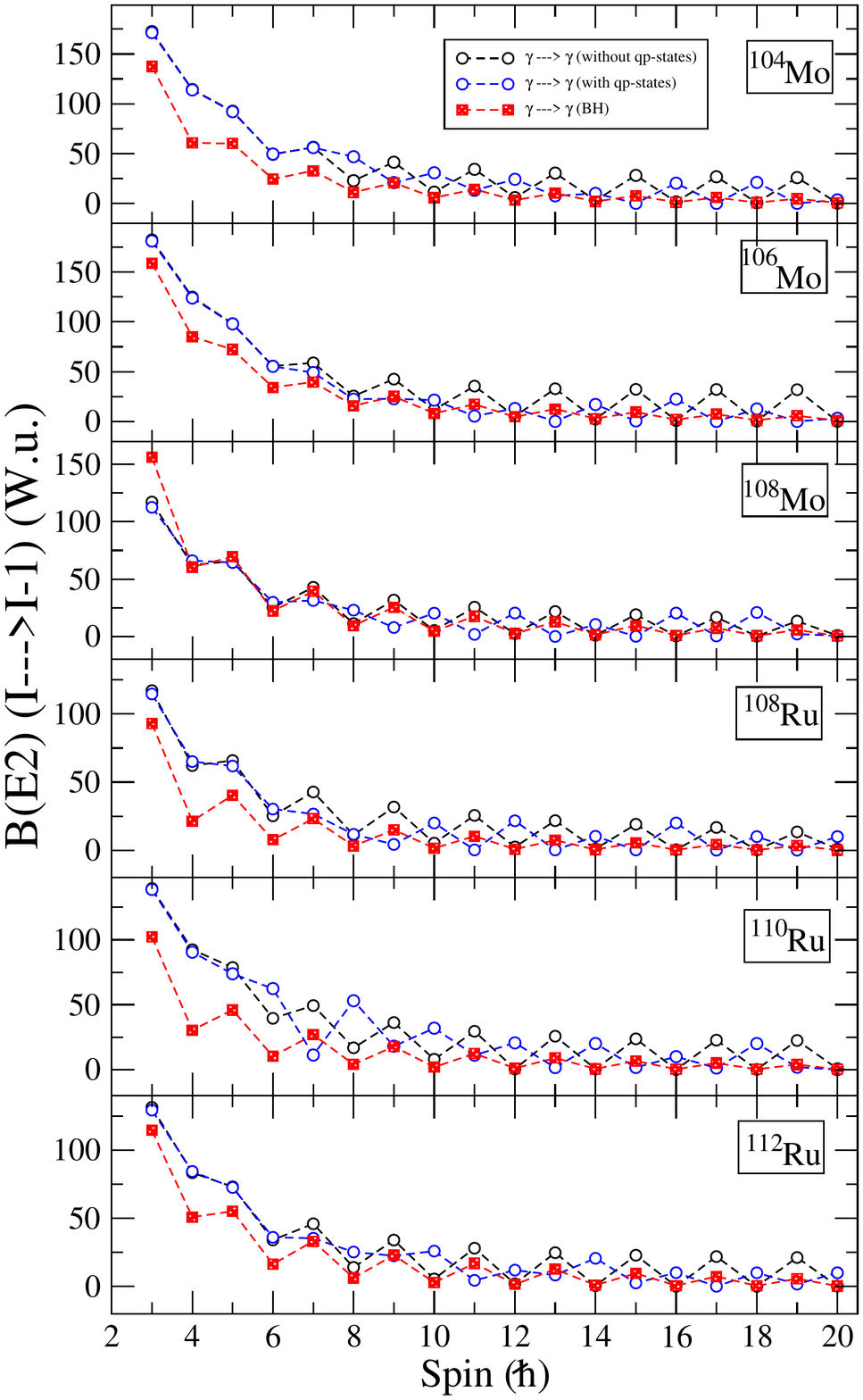}}
\caption{(Color online) Reduced E2  probabilities for $\Delta I=1$ transitions between
the $\gamma$ band members. The blue and black circles show the  TPSM values with and without 
coupling to the quasiparticle excitations, respectively.  The red squares 
show the BH values.  }\label{fig:Mo_Ru_BE2_I_I_1_g_g}
\end{figure}

\begin{figure}[h] 
 \centerline{\includegraphics[trim=0cm 0cm 0cm
0cm,width=0.5\textwidth,clip]{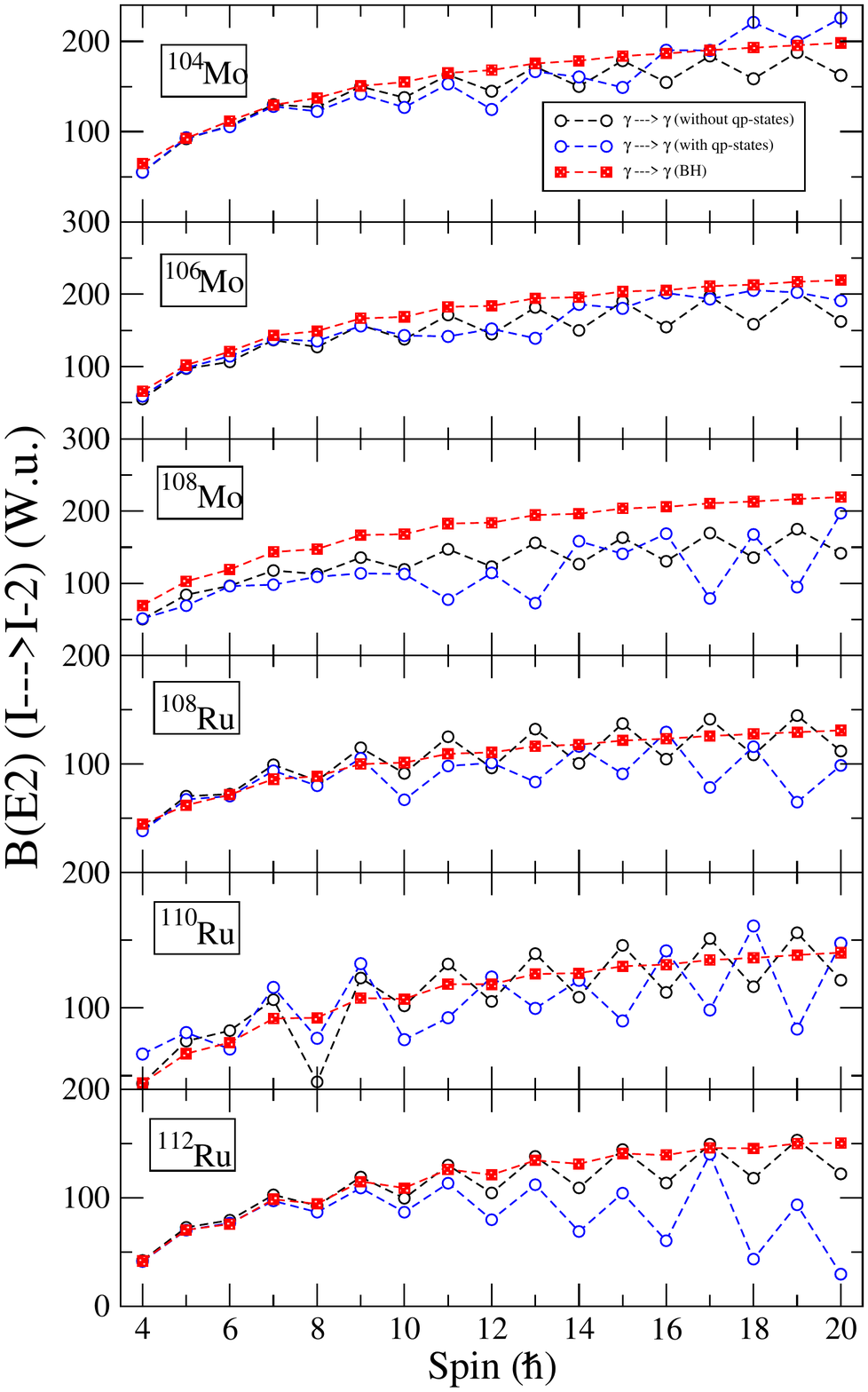}}
\caption{(Color online) Reduced E2  probabilities for $\Delta I=2$ transitions between
the $\gamma$ band members. The blue and black circles show the  TPSM values with and without 
coupling to the quasiparticle excitations, respectively.  The red squares 
show the BH values.  }\label{fig:Mo_Ru_BE2_I_I_2_g_g}
\end{figure}

\begin{figure}[h] 
 \centerline{\includegraphics[trim=0cm 0cm 0cm
0cm,width=0.45\textwidth,clip]{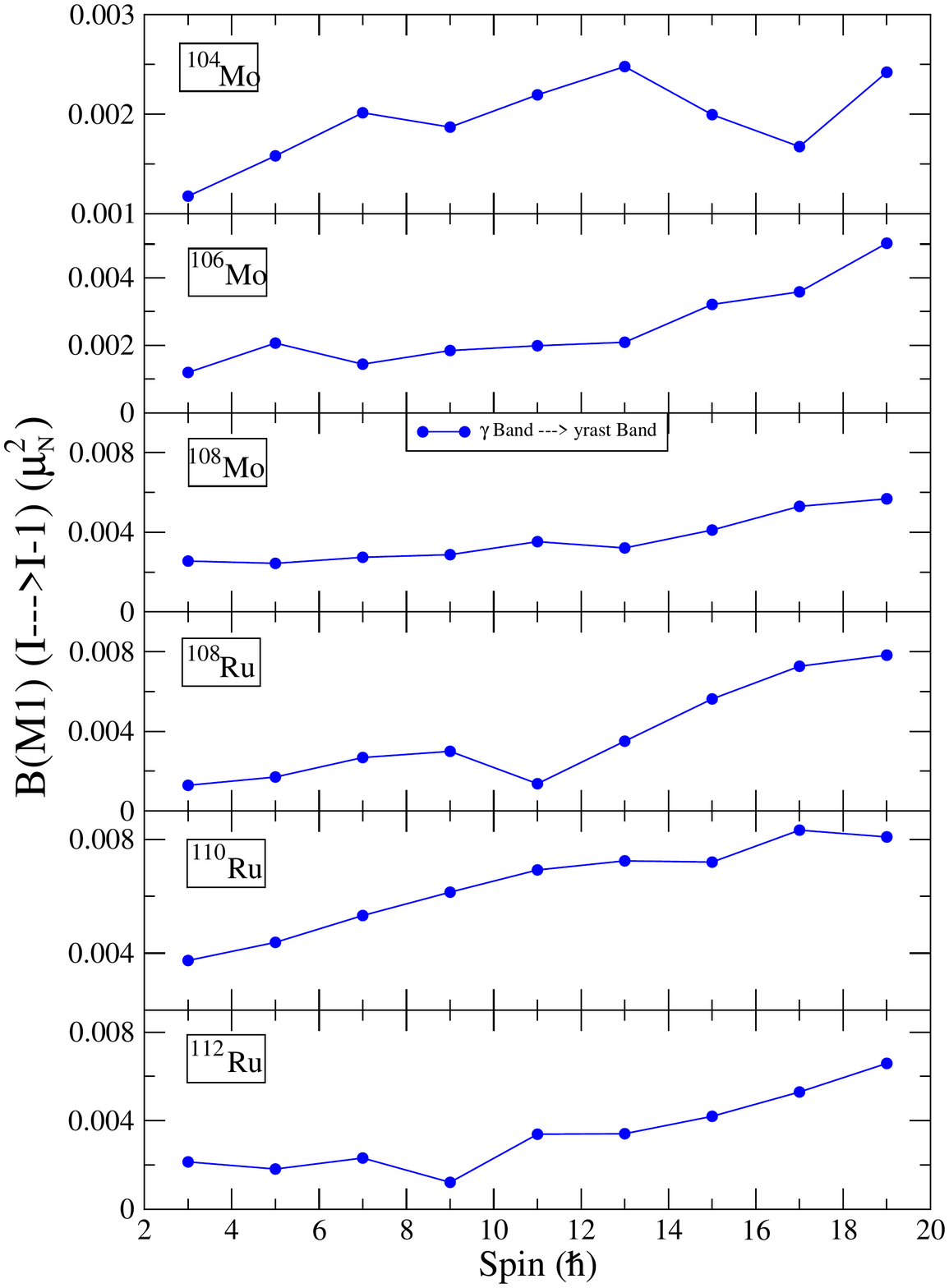}}
\caption{(Color online) Reduced M1  probabilities for $\Delta I=1$ transitions from
the $\gamma$ band to the ground band.    }\label{fig:Mo_Ru_BM1_I_I_1_g_y}
\end{figure}
\begin{figure}[h] 
 \centerline{\includegraphics[trim=0cm 0cm 0cm
0cm,width=0.45\textwidth,clip]{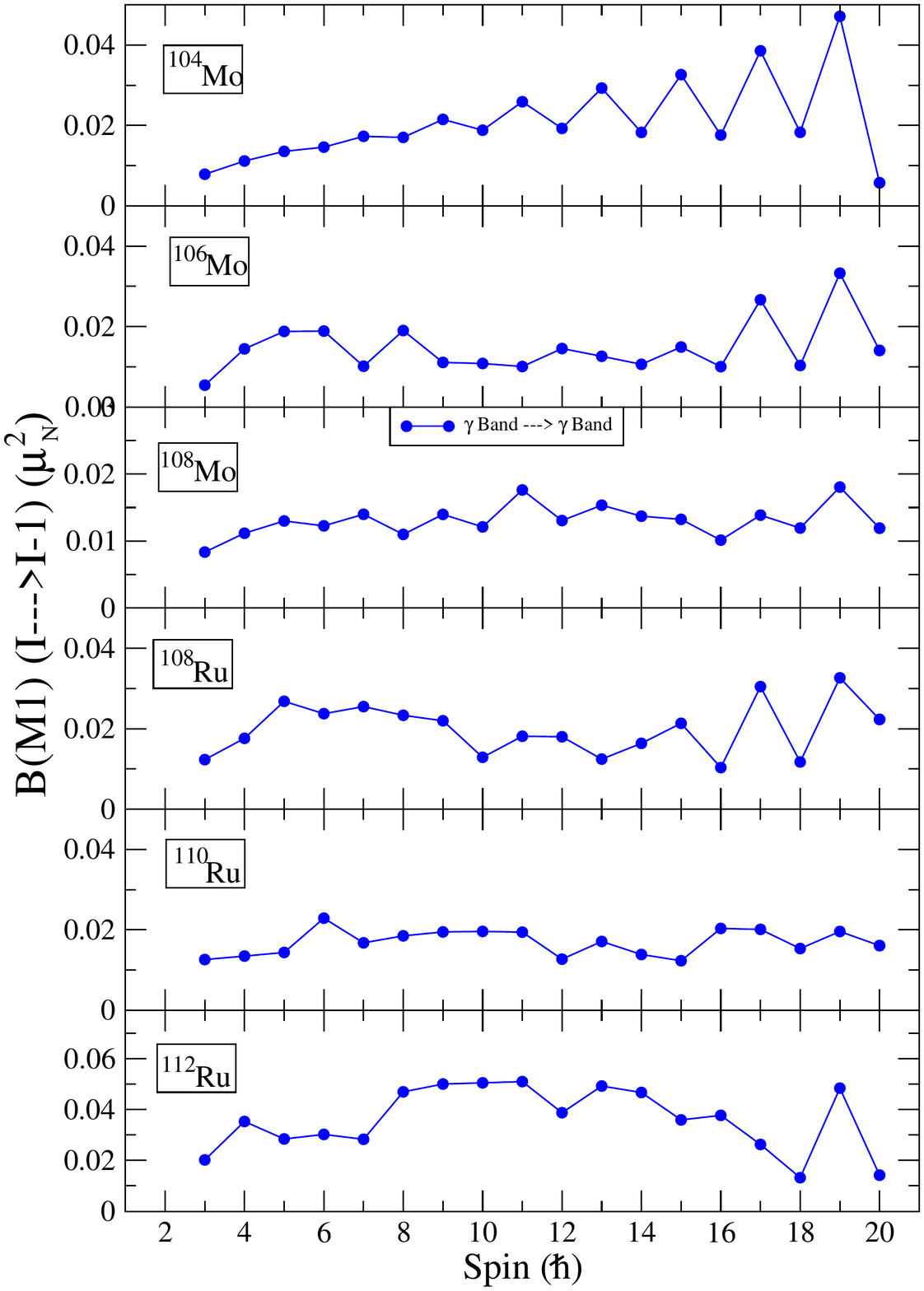}}
\caption{(Color online) Reduced M1  probabilities for $\Delta I=1$ transitions between
the $\gamma$ band members. }\label{fig:Mo_Ru_BM1_I_I_1_g_g}
\end{figure}

The fact that the TPSM results with quasiparticle excitations account
for the experimental staggering parameter is surprising, 
because the even-I-down staggering is associated with $\gamma$
softness (Wilet-Jean limit) in the context of the collective model.  
The TPSM assumes a fixed value of $\gamma$ and restricting the configuration space to the K-projections from the zero quasiparticle state only, always 
results in odd-I-low staggering. The reversal of the  staggering
phase, induced by coupling to the quasiparticle 
degrees of freedom, appears to be a different mechanism than the spread of the collective wave function.   
It is evident from Table \ref{tab1} that there is no correlation between the value of the triaxialty 
parameter $\gamma$  of the mean-field and the staggering phase
(see the Ru chain). The only visible trend is  that for  "axial" nuclei, which have a high-lying $\gamma$ band, $E(2^+_2)/E(2^+_1)\sim10$, 
the triaxiality parameter scatters around $\gamma=20^\circ$, and for "triaxial" nuclei, which have a low-lying $\gamma$ band $E(2^+_2)/E(2^+_1)<5$, 
it  scatters around $\gamma=30^\circ$. This trend is expected when one interprets the $\gamma$-band as an anharmonic tidal wave, intermediating 
the harmonic and static limits (for discussion  of tidal waves in
weakly deformed nuclei, see Ref.~\cite{FR11}.

The question that naturally arises is what is 
the amount of the quasiparticle mixing  in the even-I-down case  as
compared to the odd-I-down example. The mixing amplitudes are plotted in Figs.~\ref{fig:pRu} for
$^{108}$Ru and $^{112}$Ru as two examples belonging to former and
and latter cases, respectively. The amplitudes are shown separately for
vacuum, two- and four-quasiparticle configurations.
In $^{108}$Ru, both yrast and $\gamma$ bands are dominated by the vacuum
configuration up to I=8 and above this spin value, the
two-quasiparticle configurations dominate. This is due to the
well established crossing of the two-quasiparticle aligned band with the
ground-state band. For $^{112}$Ru, the only difference in comparison to
$^{108}$Ru is that crossover occurs at I=10 rather than at
I=8. In particular, the
magnitude of the two-quasiparticle admixture is very similar in
the two nuclei. One would have expected that due to smaller
contribution from quasiparticle excitations, $^{112}$Ru maintains the
odd-I-low staggering of the vacuum state. However, considering the
similar quasiparticle admixtures for $^{108}$Ru
and  $^{112}$Ru, the reason for having different staggering phases in
the two nuclei must be rooted in the nature of the quasineutron states at the Fermi level . 

To further probe the dependence of the $\gamma$-band staggering on the magnitude of 
triaxiality,  Fig.~\ref{fig:sRu} displays $S(I)$ for 
$^{108}$Ru and $^{112}$Ru with different values of $\epsilon'$. 
 The phase of $S(I)$ remains even-I-down, as seen at axial shape,
 for any value of $\epsilon'$  chosen in $^{108}$Ru.
 On the other hand, in $^{112}$Ru the phase of $S(I)$ changes sign with increasing triaxiality.
In order to reproduce the observed phase for the
$\gamma$-band, the triaxial deformation parameter of $\epsilon'=0.13$ ($\gamma=24^0$)
is needed. This suggests that the quasineutron states at the Fermi level couple differently to the 
triaxial potential in the two cases.

\subsection{Comparison of the transitional probabilities : TPSM and Bohr Hamiltonian}\label{sec:TPSMvsBH}

In the framework of the collective model, even-I-down staggering
pattern indicates a $\gamma$-soft potential energy
surface of the Bohr Hamiltonian,  and the odd-I-down suggests the presence of a substantial minimum at 
a finite $\gamma-$ value \cite{NV91,MC11, McCutchan07}. 
As demonstrated in the preceding section, the TPSM reproduces the experimental $S(I)$ 
values without any obvious relation to $\gamma$-softness. In an effort
to elucidate the results obtained in the TPSM approach,
we have also performed collective model
calculations by adjusting its parameters to the TPSM transition energies.
We employed a simplified version Bohr Hamiltonian (BH) of Ref. \cite{MC11} that assumes irrotational-flow inertial 
parameters and a two-parameter potential energy of the form
\begin{equation}\label{GCM3}
 {\hat H}_{GCM3}=\hat\Lambda^2+\chi\left[1-\cos3\gamma+\xi\cos^2 3\gamma\right],
  \end{equation}   
where $\hat\Lambda^2$ is the O(6) part of the Bohr kinetic energy operator, $\chi$ controls the softness and $\xi$ the depth of the minimum at finite $\gamma$.
The energy in units of $E(2^+_1)$ depend only on $\chi$ and $\xi$. The
two parameters have been fixed by a least-squares-fit to the TPSM energies of
the lowest members of the ground- and $\gamma-$ bands. The standard quadrupole operator is used, and its scale 
is adjusted to the experimental value of $B(E2,2^+_1\rightarrow 0^+_1)$.
 
To compare the results obtained in the two approaches, we have chosen
Ru- and Mo-isotopes as illustrative examples. The results for the  other
isotopic chains shall be presented in a separate detailed
comparison of the two approaches \cite{gowhar2021}. 
 In Fig. \ref{fig:MoRusTPSMvsBH}, the staggering parameter obtained in the
 two approaches is compared and it is eviden from the figure that TPSM staggering is
 reasonably reproduced by the BH model. In order to further probe the
 two approaches, we have also evaluated the transition probabilities.
 Figs. \ref{fig:Mo_Ru_BE2_I_I_2_g_y} -\ref{fig:Mo_Ru_BE2_I_I_2_g_g}
 depict the TPSM $B(E2)$ values and compares them with the values obtained  from the BH approach.  
 Fig.  \ref{fig:Mo_Ru_BE2_I_I_2_g_y} indicates that at low spin, the TPSM predicts large $B(E2)$ values for the
  transitions  $I_\gamma\rightarrow (I-2)_g$ from the $\gamma$ to the
  ground band, whereas according to the BH model only the transition
  $2^+_2\rightarrow 0^+_1$ is significantly large. All other 
 transitions are practically quenched. At high spin, both models
 predict very small $B(E2)$ for the transitions. The 
 low-spin BH values for $^{108,112}$Ru are substantially smaller than
 the corresponding TPSM values. 
   
Fig. \ref{fig:Mo_Ru_BE2_I_I_1_g_y}  shows 
the transitions
$I_\gamma\rightarrow (I-1)_g$ from $\gamma$- to the 
ground- band. The $B(E2)$ values for the $3^+_1\rightarrow 2^+_1$ are
similar for both the models in the case of the Mo isotopes, but that the low-spin BH values for $^{108,110}$Ru are substantially smaller than the TPSM ones. Similar to the  $I_\gamma\rightarrow (I-2)_g$ transitions, $B(E2, I_\gamma \rightarrow (I-1)_g$ 
 decrease more rapidly with $"I"$ for the BH and for the TPSM. TPSM
 calculations for $B(M1)$ transition probabilities, shown in
 Fig. \ref{fig:Mo_Ru_BM1_I_I_1_g_y}, indicate that transitions are almost pure $E2$.
Fig. \ref{fig:Mo_Ru_BE2_I_I_0_g_y}  shows that for the transitions  $I_\gamma\rightarrow I_g$ from the $\gamma$- to the ground-band,
the $B(E2)$ from BH are similar to the TPSM value without the quasiparticle admixture.  
 Fig. \ref{fig:Mo_Ru_BE2_I_I_1_g_g}  shows that for the transitions  $I_\gamma\rightarrow (I-1)_\gamma$ between
the members of the $\gamma$-band, the BH values show the staggering pattern of the TPSM without the quasiparticle admixtures.
The TPSM depicts the reversed pattern after the coupling to the quasiparticles is taken into account.  
At higher spins the BH values decrease more rapidly than the TPSM
values. However,   according to the TPSM calculations,
shown in Fig. \ref{fig:Mo_Ru_BM1_I_I_1_g_g}, the transitions are predominantly M1. It will be difficult to identify the differences in the staggering pattern, because the E2 part is of the order of 10-20 \%. 

The $I$ dependence of the BH results can be qualitatively understood
as follows.  The kinetic part   
 $\hat\Lambda^2$ of the BH in Eq. \ref{GCM3}  becomes dominant with increasing $I$. That is, it approaches the limit of the $\gamma$-independent  Wilets-Jean model. As discussed in detail in Ref. \cite{MC07},  
the states are grouped into SO(5) seniority muliplets ($\nu$=0,1,2,...). 
[See, e.g., Fig. 1 of Ref.  \cite{MC07}. The reader can read the label "N" in 
that figure as if it were "$\nu$"].  There is a parity quantum number 
in R$^5$, which goes as $(-1)^\nu$.  The quadrupole operator $Q$, which generates the $E2$ transitions,  carries $\nu=1$, and thus 
negative R$^5$ parity.  Between the SO(5) triangularity and parity 
constraints, this means $Q$ can only ladder $\nu$ up or down by 1. This 
hardly is a surprise, since $Q$ is the 5-dimensional quadrupole oscillator ladder 
operator, and the oscillator is a special case of SO(5) symmetry.
In summary,  the following selection rules act in the Wilets-Jean limit:\\
$I_\gamma \rightarrow (I-2)_g~~~~(I_\gamma$ even) -- forbidden by $\Delta$v=2\\
$I_\gamma \rightarrow (I-1)_g~~~~(I_\gamma$ odd ) -- forbidden by $\Delta$v=2\\
 $I_\gamma \rightarrow I_g~~~~~~~~~~~~~(I_\gamma$ even) -- allowed by $~~~\Delta$v=1\\
$I_\gamma \rightarrow (I-1)_\gamma~~~~(I_\gamma$ even) -- forbidden by $\Delta$v=0\\
$I_\gamma \rightarrow (I-1)_\gamma~~~~(I_\gamma$ odd ) -- allowed by $~~~\Delta$v=1\\

The rapid decrease of the $B(E2,I_\gamma\rightarrow (I-2)_g)$ in Fig. \ref{fig:Mo_Ru_BE2_I_I_2_g_y} and of
the $B(E2,I_\gamma\rightarrow (I-1)_g)$ in Fig. \ref{fig:Mo_Ru_BE2_I_I_1_g_y} reflects that they are forbidden 
in the Wilets-Jean limit, which is being approached with increasing $I$. The
$B(E2,I_\gamma \rightarrow I_g)$ in Fig. \ref{fig:Mo_Ru_BE2_I_I_0_g_y} remain large because they are allowed in the 
Wilets-Jean limit.
The analog holds for the $B(E2, I_\gamma \rightarrow (I-1)_\gamma)$ in Fig. \ref{fig:Mo_Ru_BE2_I_I_1_g_g}, which
generates the staggering pattern.
The transitions from the states with $I$ even are quenched because
they are forbidden and the transitons from odd $I$
are not because they are allowed as per selection rules.

To differentiate the BH and TPSM models, it seem most promising to measure the $B(E2,I_\gamma\rightarrow (I-2)_g$
for $I>2$ and $B(E2,I_\gamma\rightarrow (I-1)_g$ for $I>3$, which are much smaller for the BH than for the TPSM. 

\section{Conclusions}

In summary, a systematic investigation using the TPSM approach has been performed for 
nuclei in the Segr\'e  chart, where $\gamma$-bands, both for even- and
odd-spin signatures, are known up to quite high spin.
When only the vacuum configuration is taken into account, 
 the $\gamma$-band staggering parameter, $S(I)$ has always odd-I-down.
It has been shown that
for almost all the nuclei,  the phase of $S(I)$
changes  from odd-I-down to even-I-down,  when
 the quasiparticle excitations are taken into account in the
 configuration mixing framework. The only 
exceptions are  the four nuclei of $^{76}$Ge, $^{112}$Ru, $^{170}$Er and $^{232}$Th. The staggering parameter 
calculated using the TPSM approach reproduces quite well the
corresponding experimental values.
In particular,  the odd-I-down pattern, observed only in the
aforementioned four nuclei nuclei, is reproduced in the TPSM approach.
 
The traditional interpretation of the collective model based on Bohr Hamiltonian,
 associates the even-I-down pattern with a $\gamma$-soft shape and the 
 odd-I-down pattern with the presence of static triaxiality.  In
 particular, $^{76}$Ge and $^{112}$Ru
 have been presented as rare examples of rigid triaxiality
 \cite{McCutchan07,GH14,YT13,110rudata}.  In contrast, TPSM 
 successfully accounts for the "$\gamma$-soft" staggering pattern
by assuming a rigid triaxial shape and explicitly considering the
 quasiparticle excitations that are expected to drive the system
 to the vibrational mode. However, it has been noted that
 the magnitude of the quasiparticle content in $^{112}$Ru
 ($\gamma$-rigid case) and $^{108}$Ru ($\gamma$-soft case)  wavefunctions is similar.
This suggests that the  reason for staggering phase reversal obtained for
all the studied nuclei, except for $^{76}$Ge, $^{112}$Ru, $^{170}$Er
and $^{232}$Th is rooted in the nuclear shell structure.

To further  examine the results obtained in the TPSM approach, we have
also solved Bohr Hamiltonain. 
We studied staggering and $B(E2)$  transition
  probabilities predicted by a two-parameter version of the Bohr Hamiltonian. The two parameters, 
  which determine the softness and the presence of a minimum at finite $"\gamma"$ of the collective potential,
  were adjusted to the energies of the lowest members of the ground- and
  $\gamma$-bands as obtained in the TPSM approach.
  The fit reproduces quite well  the TPSM staggering pattern.
  However, there are significant differences  between  the
  reduced transition probabilities, which may allow one to delineate
  the two approaches. In particular,    
  the Bohr Hamiltonian values of $B(E2, I_\gamma \rightarrow(I-2)_g)$ and $B(E2, I_\gamma \rightarrow(I-1)_g)$ 
  for the transition from the $"\gamma"$ to the ground band fall off
  rapidly for $I>4 (3) $, 
  whereas the TPSM values decrease much slower with $I$.  The
  expermental data on transition probabilities is needed to shed light
  on the nature of the predicted transitions.
  
\section{Acknowledgements}
The authors thank M. Caprio for illuminating discussions concerning
the SO(5) selection rules. The work was partly supported by the US DoE grant DE-FG02-95-ER4093. 
Two of us (GHB and JAS) would like to
acknowledge Department of Science and Technology, Govt. of India for providing financial support under Project no. CRG/2019/004960
to carry out a part of the research work.

\end{document}